\documentclass[sigconf]{acmart}

\AtBeginDocument{%
  \providecommand\BibTeX{{%
    \normalfont B\kern-0.5em{\scshape i\kern-0.25em b}\kern-0.8em\TeX}}}

\copyrightyear{2021}
\acmYear{2021}
\setcopyright{rightsretained}
\acmConference[CHI '21]{CHI Conference on Human Factors in Computing Systems}{May 8--13, 2021}{Yokohama, Japan}
\acmBooktitle{CHI Conference on Human Factors in Computing Systems (CHI '21), May 8--13, 2021, Yokohama, Japan}
\acmDOI{10.1145/3411764.3445049}
\acmISBN{978-1-4503-8096-6/21/05}

\usepackage{tabularx}
\usepackage{amsmath}
\usepackage{balance}

\newcommand{\tlhighlight}[1]{#1}

\newcommand{\tlcomment}[1]{}

\newcommand{\lpcomment}[1]{}

\begin{document}

\title{Screen2Vec: Semantic Embedding of GUI Screens and GUI Components}

\author{Toby Jia-Jun Li}
\authornote{Both authors contributed equally.}
\email{tobyli@cs.cmu.edu}
\affiliation{%
  \institution{Carnegie Mellon University}
  \city{Pittsburgh}
  \state{PA}
}

\author{Lindsay Popowski}
\authornotemark[1]
\email{lpopowski@g.hmc.edu}
\affiliation{%
  \institution{Harvey Mudd College}
  \city{Claremont}
  \state{CA}
}

\author{Tom M. Mitchell}
\email{tom.mitchell@cs.cmu.edu}
\affiliation{%
  \institution{Carnegie Mellon University}
  \city{Pittsburgh}
  \state{PA}
}

\author{Brad A. Myers}
\email{bam@cs.cmu.edu}
\affiliation{%
  \institution{Carnegie Mellon University}
  \city{Pittsburgh}
  \state{PA}
}


\begin{abstract}
Representing the semantics of GUI screens and components is crucial to data-driven computational methods for modeling user-GUI interactions and mining GUI designs. Existing GUI semantic representations are limited to encoding either the textual content, the visual design and layout patterns, or the app contexts. Many representation techniques also require significant manual data annotation efforts. This paper presents \texttt{Screen2Vec}, a new self-supervised technique for generating representations in embedding vectors of GUI screens and components that encode all of the above GUI features without requiring manual annotation using the context of user interaction traces. \texttt{Screen2Vec} is inspired by the word embedding method \texttt{Word2Vec}, but uses a new two-layer pipeline informed by the structure of GUIs and interaction traces and incorporates screen- and app-specific metadata. Through several sample downstream tasks, we demonstrate \texttt{Screen2Vec}'s key useful properties: representing between-screen similarity through nearest neighbors, composability, and capability to represent user tasks.

\end{abstract}




\maketitle

\section{Introduction}
With the rise of data-driven computational methods for modeling user interactions with graphical user interfaces (GUIs), the GUI screens have become not only interfaces for human users to interact with the underlying computing services, but also valuable data sources that encode the underlying task flow, the supported user interactions, and the design patterns of the corresponding apps, which have proven useful for AI-powered applications. For example, programming-by-demonstration (PBD) intelligent agents such as~\cite{li_sugilite:_2017,li_pumice:_2019,sereshkeh2020vasta} use task-relevant entities and hierarchical structures extracted from GUIs to parameterize, disambiguate, and handle errors in user-demonstrated task automation scripts. \textsc{Erica}~\cite{deka_erica:_2016} mines a large repository of mobile app GUIs to enable user interface (UI) designers to search for example design patterns to inform their own design. Kite~\cite{li_kite:_2018} extracts task flows from mobile app GUIs to bootstrap conversational agents. 


Semantic representations of GUI screens and components, where each screen and component is encoded as a vector (known as the \textit{embedding}), are highly useful in these applications. The representations of GUI screens and components can be used to also represent other entities of interest. For example, a task in an app can be modeled as a sequence of GUI actions, where each action can be represented as a GUI screen, a type of interaction (e.g., click), and the component that is interacted with on the screen. An app can be modeled as a collection of all its screens, or a large collection of user interaction traces of using the app. Voice shortcuts in mobile app deep links~\cite{azim_ulink:_2016} can be modeled as matching the user's intent expressed in natural language to the target GUI screens. The representation of the screen that the user is viewing or has previously viewed can also be used as the context to help infer the user's intents and activities in predictive intelligent interfaces. The semantic embedding approach represents GUI screens and components in a \textit{distributed} form~\cite{bengio2009learning} (i.e., an item is represented across multiple dimensions) as continuous-valued vectors, making it especially suitable for use in popular machine learning models.

However, existing approaches of representing GUI screens and components are limited. One type of approach solely focuses on capturing the text on the screen, treating the screen as a bag of words or phrases. For example, \textsc{Sugilite}~\cite{li_sugilite:_2017} uses exact matches of text labels on the screen to generalize the user demonstrated tasks. \textsc{Sovite}~\cite{li_sovite:_2020} uses the average of individual word embedding vectors for all the text labels on the screen to represent the screen for retrieving relevant task intents. This approach can capture the semantics of the screen's textual content, but misses out on using the information encoded in the layout and the design pattern of the screen and the task context encoded in the interactivity and meta-data of the screen components.

Another type of approach focuses on the visual design patterns and GUI layouts. \textsc{Erica}~\cite{deka_erica:_2016} uses an unsupervised clustering method to create semantic clusters of visually similar GUI components. Liu et al.'s approach~\cite{liu2018learning} leverages the hierarchical GUI structures, the class names of GUI components, and the visual classifications of graphical icons to annotate the design semantics of GUIs. This type of approach has been shown to be able to determine the category of a GUI component (e.g., list items, tab labels, navigation buttons), the ``UX concept'' semantics of buttons (e.g., ``back'', ``delete'', ``save'', and ``share''), and the overall type of task flow of screens (e.g., ``searching'', ``promoting'', and ``onboarding''). However, it does not capture the content in the GUIs---two structurally and visually similar screens with different content (e.g., the search results screen in a restaurant app and a hotel booking app) will yield similar results. 

There have been prior approaches that combine the textual content and the visual design patterns~\cite{pasupat-etal-2018-mapping,li_mapping:_2020}. However, these approaches use supervised learning with large datasets for very specific task objectives. Therefore they require significant task-specific manual data labeling efforts, and their resulting models cannot be used in different downstream tasks. For example, Pasupat et al.~\cite{pasupat-etal-2018-mapping} create a embedding-based model that can map the user's natural language commands to web GUI elements based on the text content, attributes, and spatial context of the GUI elements. Li et al.'s work~\cite{li_mapping:_2020} describes a model that predicts sequences of mobile GUI action sequences based on step-by-step natural language descriptions of actions. Both models are trained using large manually-annotated corpora of natural language utterances and the corresponding GUI actions.

We present a new \textit{self-supervised} technique (i.e., the type of machine learning approach that trains a model without human-labeled data by withholding some part of the data, and tasking the network with predicting it) \texttt{Screen2Vec} for generating more comprehensive semantic representations of GUI screens and components. \texttt{Screen2Vec} uses the screens' textual content, visual design and layout patterns, and app context meta-data. \texttt{Screen2Vec}'s approach is inspired by the popular word embedding method \texttt{Word2Vec}~\cite{mikolov_distributed_2013}, where the embedding vector representations of GUI screens and components are generated through the process of training a prediction model. However, unlike \texttt{Word2Vec}, \texttt{Screen2Vec} uses a two-layer pipeline informed by the structures of GUIs and GUI interaction traces and incorporates screen- and app-specific metadata. 

The embedding vector representations produced by \texttt{Screen2Vec} can be used in a variety of useful downstream tasks such as nearest neighbor retrieval, composability-based retrieval, and representing mobile tasks. The self-supervised nature of \texttt{Screen2Vec} allows its model to be trained without any manual data labeling efforts---it can be trained with a large collection of GUI screens and the user interaction traces on these screens such as the \textsc{Rico}~\cite{deka_rico:_2017} dataset.

Along with this paper, we also release the open-source\footnote{A pre-trained model and the \texttt{Screen2Vec} source code are available at: \tlhighlight{\url{https://github.com/tobyli/screen2vec}}} code of \texttt{Screen2Vec} as well as a pre-computed \texttt{Screen2Vec} model trained on the \textsc{Rico} dataset~\cite{deka_rico:_2017} (more in Section~\ref{sec:dataset}). The pre-computed model can encode the GUI screens of Android apps into embedding vectors off-the-shelf. The open-source code can be used to train models for other platforms given the appropriate dataset of user interaction traces.

\texttt{Screen2Vec} addresses an important gap in prior work about computational HCI research. The lack of comprehensive semantic representations of GUI screens and components has been identified as a major limitation in prior work in GUI-based interactive task learning (e.g.,~\cite{li_pumice:_2019, sereshkeh2020vasta}), intelligent suggestive interfaces (e.g.,~\cite{Chen2019MessageOnTap}), assistive tools (e.g.,~\cite{bigham2009trailblazer}), and GUI design aids (e.g.,~\cite{swearngin2018rewire,lee2020guicomp}). \texttt{Screen2Vec} embeddings can encode the semantics, contexts, layouts, and patterns of GUIs, providing representations of these types of information in a form that can be easily and effectively incorporated into popular modern machine learning models.

This paper makes the following contributions: 

\begin{enumerate}

\item \texttt{Screen2Vec}: a new self-supervised technique for generating more comprehensive semantic embeddings of GUI screens and components using their textual content, visual design and layout patterns, and app meta-data.

\item An open-sourced GUI embedding model trained using the \texttt{Screen2Vec} technique on the \textsc{Rico}~\cite{deka_rico:_2017} dataset that can be used off-the-shelf. 

\item Several sample downstream tasks that showcase the model's usefulness.

\end{enumerate}

\section{Our Approach}
\label{sec:approach}

\begin{figure*}
	\includegraphics[width=\textwidth]{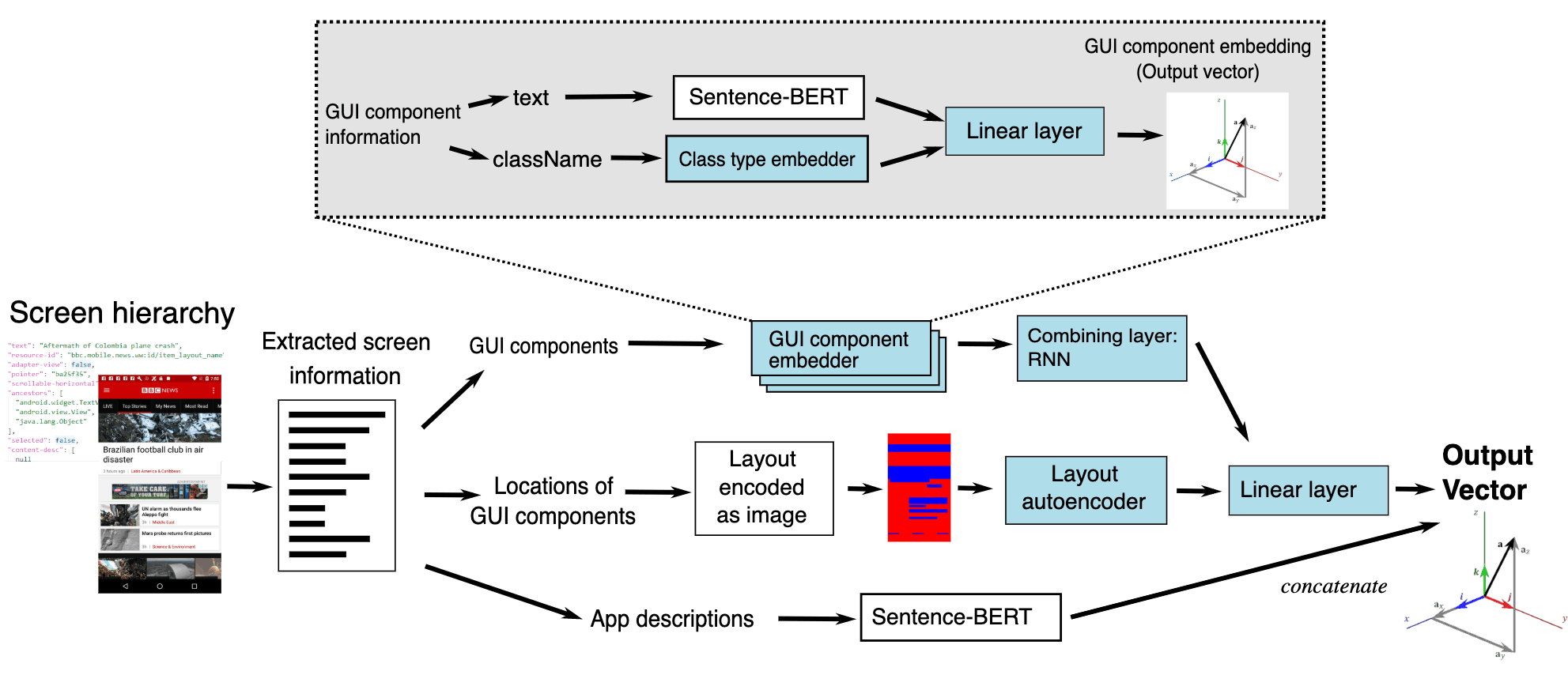}
    \Description{The figure showing the architecture of the Screen2Vec model. The pipeline of Screen2Vec consists of two levels: the GUI component level and the GUI screen level.}
	\caption{The two-level architecture of \texttt{Screen2Vec} for generating GUI component and screen embeddings. The weights for the steps in teal color are optimized during the training process.}
	\label{fig:architecture}
\end{figure*}

Figure~\ref{fig:architecture} illustrates the architecture of \texttt{Screen2Vec}. Overall, the pipeline of \texttt{Screen2Vec} consists of two levels: the GUI component level (shown in the gray shade) and the GUI screen level. We will first describe the approach at a high-level here, and then explain the details in Section~\ref{sec:models}.

The GUI component level model encodes the textual content and the class type of a GUI component into a 768-dimensional\footnote{We decided to produce 768-dimensional vectors so that they can be directly used with the 768-dimensional vectors produced by the pre-trained Sentence-BERT model with its default settings~\cite{reimers-2019-sentence-bert}} embedding vector to represent the GUI component (e.g., a button, a textbox, a list entry etc.). This GUI component embedding vector is computed with two inputs: (1) a 768-dimensional embedding vector of the text label of the GUI component, encoded using a pre-trained Sentence-BERT~\cite{reimers-2019-sentence-bert} model; and (2) a 6-dimensional class embedding vector that represents the class type of the GUI component, which we will discuss in detail later in Section~\ref{sec:models}. The two embedding vectors are combined using a linear layer, resulting in the 768-dimensional GUI component embedding vector that represents the GUI component. The class embeddings in the class type embedder and the weights in the linear layer are optimized through training a Continuous Bag-of-Words (CBOW) prediction task: for each GUI component on each screen, the task predicts the current GUI component using its context (i.e., all the other GUI components on the same screen). The training process optimizes the weights in the class embeddings and the weights in the linear layer for combining the text embedding and the class embedding.

The GUI screen level model encodes the textual content, visual design and layout patterns, and app context of a GUI screen into an 1536-dimensional embedding vector. This GUI screen embedding vector is computed using three inputs: (1) the collection of the GUI component embedding vectors for all the GUI components on the screen (as described in the last paragraph), combined into a 768-dimension vector using a recurrent neural network model (RNN), which we will discuss more in Section~\ref{sec:models}; (2) a 64-dimensional layout embedding vector that encodes the screen's visual layout (details later in Section~\ref{sec:models}); and (3) a 768-dimensional embedding vector of the textual App Store description for the underlying app, encoded with a pre-trained Sentence-BERT~\cite{reimers-2019-sentence-bert} model. These GUI and layout vectors are combined using a linear layer, resulting in a 768-dimensional vector. After training, the description embedding vector is concatenated on, resulting in the 1536-dimensional GUI screen embedding vector (if included in the training, the description dominates the entire embedding, overshadowing information specific to that screen within the app). \tlhighlight{The weights in the RNN layer for combining GUI component embeddings and the weights in the linear layer for producing the final output vector are similarly trained on a CBOW prediction task on a large number of interaction traces (each represented as a sequence of screens). For each trace, a sliding window moves over the sequence of screens. The model tries to use the representation of the context (the surrounding screens) to predict the screen in the middle. See Section~\ref{sec:models} for more details.}

However, unlike the GUI component level embedding model, the GUI screen level model is trained on a screen prediction task in the user interaction traces of using the apps. Within each trace, the training task tries to predict the current screen using other screens in the same trace. 


\subsection{Dataset}
\label{sec:dataset}
We trained \texttt{Screen2Vec} on the open-sourced \textsc{Rico}\footnote{Available at: \url{http://interactionmining.org/rico}} dataset~\cite{deka_rico:_2017}. The \textsc{Rico} dataset contains interaction traces on 66,261 unique GUI screens from 9,384 free Android apps collected using a hybrid crowdsourcing plus automated discovery approach. For each GUI screen, the \textsc{Rico} dataset includes a screenshot image (that we did not use in \texttt{Screen2Vec}), and the screen's ``view hierarchy'' in a JSON file. The view hierarchy is structurally similar to a DOM tree in HTML; it  starts with a root view, and contains all its descents in a tree. The node for each view includes the class type of this GUI component, its textual content (if any), its location as the bounding box on the screen, and various other properties such as whether it is clickable, focused, or scrollable, etc. Each interaction trace is represented as a sequence of GUI screens, as well as information about which (x, y) screen location was clicked or swiped on to transit from the previous screen to the current screen.  

\subsection{Models}
\label{sec:models}
This section explains the implementation details of each key step in the pipeline shown in Figure~\ref{fig:architecture}.
\paragraph{GUI Class Type Embeddings}

To represent the class types of GUI components, we trained a class embedder to encode the class types into the vector space. We used a total of 26 class categories: the 22 categories that were present in \cite{liu2018learning}, one layout category, list and drawer categories, and an ``Other'' category. We classified the GUI component classes based on the classes of their \texttt{className} properties and, sometimes, other simple heuristic rules (see Table~\ref{tab:class_categories}). For example, if a GUI component is an instance of \texttt{EditText} (i.e., its \texttt{className} property is either \texttt{EditText}, or a class that inherits \texttt{EditText}), then it is classified as an Input. There are two exceptions: the Drawer and the List Item categories look at the \texttt{className} of the parent of the current GUI component instead of the \texttt{className} of itself. A standard PyTorch embedder (\texttt{torch.nn.Embedding}\footnote{\url{https://pytorch.org/docs/stable/generated/torch.nn.Embedding.html}}) maps each of these 26 discrete categories into a continuous 6-dimensional vector. The embedding vector value for each category is optimized during the training process for the GUI component prediction tasks so that GUI components categories that are semantically similar to each other are closer together in the vector space.

\begin{table*}[htbp]
\small
\renewcommand{\arraystretch}{1.1}
\newcolumntype{L}{>{\raggedright\arraybackslash}X}
\begin{minipage}{\textwidth}
\begin{tabularx}{\textwidth}{|l|L||l|L|}
\hline
GUI Component & Associated Class Type & GUI Component & Associated Class Type   \\
\hline
Advertisement & \texttt{AdView, HtmlBannerWebView, AdContainer} & Layouts & \texttt{LinearLayout, AppBarLayout, FrameLayout, RelativeLayout, TableLayout}  \\
Bottom Navigation & \texttt{BottomTabGroupView, BottomBar} & Button Bar & \texttt{ButtonBar}   \\
Card & \texttt{CardView} & CheckBox & \texttt{CheckBox, CheckedTextView}  \\
Drawer (Parent) & \texttt{DrawyerLayout} & Date Picker & \texttt{DatePicker}  \\
Image & \texttt{ImageView} & Image Button & \texttt{ImageButton, GlyphView, AppCompatButton, AppCompatImageButton, ActionMenuItemView, ActionMenuItemPresenter}  \\
Input & \texttt{EditText, SearchBoxView, AppCompatAutoCompleteTextView, TextView}\footnote{The property \texttt{editable} needs to be \texttt{TRUE}.} & List Item (Parent) & \texttt{ListView, RecyclerView, ListPopupWindow, TabItem, GridView}  \\
Map View & \texttt{MapView} & Multi-Tab & \texttt{SlidingTab}  \\
Number Stepper & \texttt{NumberPicker} & On/Off Switch & \texttt{Switch}  \\
Pager Indicator & \texttt{ViewPagerIndicatorDots, PageIndicator, CircileIndicator, PagerIndicator} & RadioButton & \texttt{RadioButton, CheckedTextView}  \\ 
Slider & \texttt{SeekBar} & TextButton & \texttt{Button\footnote{The GUI component needs to have a non-empty \texttt{text} property.}, TextView\footnote{The property \texttt{clickable} needs to be \texttt{TRUE}.}}  \\
Tool Bar & \texttt{ToolBar, TitleBar, ActionBar} & Video & \texttt{VideoView}  \\
Web View & \texttt{WebView} & Drawer Item & \texttt{Others category and ancestor is Drawer(Parent)} \\ 
List Item & \texttt{Others category and ancestor is List(Parent)} & Others & \texttt{...} \\
\hline
\end{tabularx}
\end{minipage}

\vspace{0.3cm}
\caption{The 26 categories (including the ``Others'' category) of GUI class types we used in \texttt{Screen2Vec} and their associated base class names. Some categories have additional heuristics, as shown in the notes. This categorization is adapted from~\cite{liu2018learning}.}
\label{tab:class_categories}
\end{table*}


 \tlcomment{TODO: add a figure showing the embeddings of GUI class types on a 2d plane}

\paragraph{GUI Component Context}
As discussed earlier, \texttt{Screen2Vec} uses a Continuous Bag-of-Words (CBOW) prediction task~\cite{mikolov_distributed_2013} for training the weights in the model, where for each GUI component, the model tries to predict it using its context. In \texttt{Screen2Vec}, we define the context of a GUI component as its 16 nearest components. The size 16 is chosen to balance the model performance and the computational cost. 

\tlhighlight{Inspired by prior work on the correlation between the semantic relatedness of entities and the spatial distance between them~\cite{li_leveraging_2014}. We tried using two different measures of screen distance for determining GUI component context in our model: \texttt{EUCLIDEAN}, which is the straight-line minimal distance on the screen (measured in pixels) between the bounding boxes of the two GUI components; and \texttt{HIERARCHICAL}, which is the distance between the two GUI components on the hierarchical GUI view tree. For example, a GUI component has a distance of 1 to its parent and children and a distance of 2 to its direct siblings.} 

\paragraph{Linear Layers}

At the end of each of the two levels in the pipeline, a linear layer is used to combine multiple vectors and shrink the combined vector into a lower-dimension vector that contains the relevant semantic content of each input. For example, in the GUI component embedding process, the model first concatenates the 768-dimensional text embedding with the 6-dimensional class embedding. The linear layer then shrinks the GUI component embedding back down to 768 dimensions. The linear layer works by creating $774 \times 768$ weights: one per pair of input dimension and output dimension. These weights are optimized along with other parameters during the training process, so as to minimize the overall total loss (loss function detail in Section~\ref{sec:training}).

In the screen embedding process, a linear layer is similarly used for combining the 768-dimensional layout embedding vector with the 64-dimensional GUI content embedding vector to produce a new 768-dimensional embedding vector that encodes both the screen content and the screen layout.

\paragraph{Text Embeddings}

We use a pre-trained Sentence-BERT language model~\cite{reimers-2019-sentence-bert} to encode the text labels on each GUI component and the Google Play store description for each app into 768-dimensional embedding vectors. This Sentence-BERT model, which is a modified BERT network~\cite{devlin_2018_bert}, was pre-trained on the SNLI~\cite{bowman_large:_2015} dataset and the Multi-Genre NLI~\cite{williams_broad:_2018} dataset with a mean-pooling strategy, as described in~\cite{reimers-2019-sentence-bert}. This pre-trained model has been shown to perform well in deriving semantically meaningful sentence and phrase embeddings where semantically similar sentences and phrases are close to each other in the vector space~\cite{reimers-2019-sentence-bert}.

\begin{figure*}
	\centering
	\includegraphics[width=\textwidth]{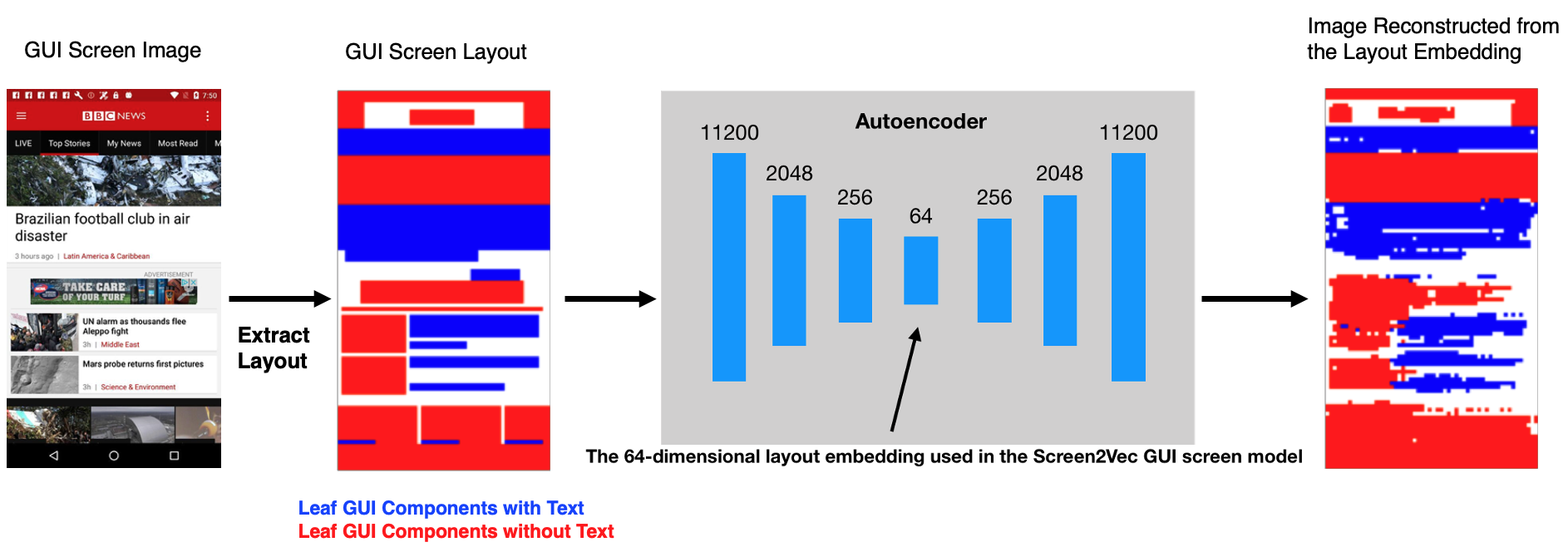}
    \Description{Screen2Vec's autoencoder can transform the screenshot of an app into an image that represents the layout of the screen}
	\caption{\texttt{Screen2Vec} extracts the layout of a GUI screen as a bitmap, and encodes this bitmap into a 64-dimensional vector using a standard autoencoder architecture where the autoencoder is trained on the loss of the output of the decoder~\cite{deka_rico:_2017}.}
	\label{fig:autoencoder}
\end{figure*}

\paragraph{Layout Embeddings}


Another important step in the pipeline is to encode the visual layout pattern of each screen. We use the layout embedding technique from~\cite{deka_rico:_2017}, where we first extract the layout of a screen from its screenshot using the bounding boxes of all the leaf GUI components in the hierarchical GUI tree, differentiating between text and non-text GUI components using different colors (Figure~\ref{fig:autoencoder}). This layout image represents the layout of the GUI screen while abstracting away its content and visual specifics. We then use an image \textit{autoencoder} to encode each image into a 64-dimensional embedding vector. The autoencoder is trained using a typical encoder-decoder architecture, that is, the weights of the network are optimized to produce the 64-dimensional vector from the original input image that can produce the best reconstructed image when decoded.

The encoder has input dimension of 11,200, and then two hidden layers of size 2,048 and 256, with output dimension of size 64; this means three linear layers of sizes $11,200 \rightarrow 2,048, 2,048 \rightarrow 256$, and $256 \rightarrow 64$. These layers have the Rectified Linear Unit (ReLU)~\cite{nair_relu_2010} applied, so the output of each linear layer is put through an activation function which transforms any negative input to 0. The decoder has the reverse architecture (three linear layers with ReLU $64 \rightarrow 256, 256 \rightarrow 2,048$, and $2,048 \rightarrow 11,200$). The layout autoencoder is trained on the process of reconstructing the input image when it is run through the encoder and the decoder; the loss is determined by the mean squared error (MSE) between the input of the encoder and the output of the decoder.   

\paragraph{GUI Embedding Combining Layer}

To combine the embedding vectors of multiple GUI components on a screen into a single fixed-length embedding vector, we use an Recurrent Neural Network (RNN): The RNN operates similarly to the linear layer mentioned earlier, except it deals with sequential data (thus the ``recurrent'' in the name). The RNN we used was a sequence of linear layers with the additional input of a \textit{hidden state}. \tlhighlight{The GUI component embeddings are fed into the RNN in the pre-order traversal order of the GUI hierarchy tree.} For the first input of GUI component embedding, the hidden state was all zeros, but for the second input, the output from the first serves as the hidden state, and so on, so that the $n^{th}$ input is fed into a linear layer along with $(n-1)^{th}$ output. The overall output is the output for the final GUI component in the sequence, which encodes parts of all of the GUI components, since the hidden states could pass on that information. This allows screens with different numbers of GUI components to have vector representations that both take all GUI components into account \emph{and} are of the same size. This RNN is trained along with all other parameters in the screen embedding model, optimizing for the loss function (detail in Section~\ref{sec:training}) in the GUI screen prediction task.


\tlcomment{report the results of this baseline?}


\subsection{Training Configurations}
\label{sec:training}

In the training process, we use 90\% of the data for training and save the other 10\% for validation. The models are trained on a cross entropy loss function with an Adam optimizer~\cite{kingma_adam_2015}, which is an adaptive learning gradient-based optimization algorithm of stochastic objective functions. For both stages, we use an initial learning rate of 0.001 and a batch size of 256.

The GUI component embedding model takes about 120 epochs to train, while the GUI screen embedding model takes 80--120 epochs depending on which version is being trained\footnote{The version without spatial information takes 80 epochs; and the one with spatial information takes 120.}. A virtual machine with 2 NVIDIA Tesla K80 GPUs can train the GUI component embedding model in about 72 hours, and train the GUI screen embedding model in about 6-8 hours.

We used PyTorch's implementation of the \texttt{CrossEntropyLoss} function\footnote{\url{https://pytorch.org/docs/stable/generated/torch.nn.CrossEntropyLoss.html}} to calculate the prediction loss. The \texttt{CrossEntropyLoss} function combines negative log likelihood loss (NLL Loss) with the log softmax function:

\begin{align*}
    CrossEntropyLoss(x,class) &= NLL\_Loss(logSoftmax(x), class))  \\
    &= -log( \frac{exp(x[class])}{\sum\nolimits_{c} exp(x[c])}) \\
    &= -x[class] + log\sum\nolimits_{c} exp(x[c])
\end{align*}

In the case of the GUI component embedding model, the total loss is the sum of the cross entropy loss for the text prediction and the cross entropy loss for the class type prediction. In calculating the cross entropy loss, each text prediction was compared to every possible text embedding in the vocabulary, and each class prediction was compared to all possible class embeddings.

In the case of the GUI screen  embedding model, the loss is exclusively for screen predictions. However, the vector $x$ does not contain the similarity between the correct prediction and \textit{every} screen in the dataset. \tlhighlight{Instead we use \textit{negative sampling}~\cite{mikolov_distributed_2013, mikolov_efficient_2013} so that we do not have to recalculate and update every screen's embedding on every training iteration, which is computationally expensive and prone to over-fitting. In each iteration, the prediction is compared to the correct screen and a \textit{sample of negative data} that consists of: a random sampling of size 128 of other screens, the other screens in the batch, and the screens in the same trace as the correct screen, used in the prediction task.} We specifically include the screens in the same trace to promote screen-specific learning in this process: This way, we can disincentive screen embeddings that are based solely on the app\footnote{Since the next screen is always within the same app and therefore shares an app description embedding, the prediction task favors having information about the specific app (i.e., app store description embedding) dominate the embedding.}, and emphasize having the model learn to differentiate the different screens within the same app. 

\subsection{Baselines}
We compared \texttt{Screen2Vec} to the following three baseline models:

\paragraph{Text Embedding Only} The \texttt{TextOnly} model replicates the screen embedding method used in \textsc{Sovite}~\cite{li_sovite:_2020}. It only looks at the textual content on the screen: the screen embedding vector is computed by averaging the text embedding vectors for all the text found on the screen. The pre-trained Sentence-BERT model~\cite{reimers-2019-sentence-bert} calculates the text embedding vector for each text. With the the \texttt{TextOnly} model, screens with semantically similar textual contexts will have similar embedding vectors.

\paragraph{Layout Embedding Only} The \texttt{LayoutOnly} model replicates the screen embedding method used in the original \textsc{Rico} paper~\cite{deka_rico:_2017}. It only looks at the visual layout of the screen: It uses the layout embedding vector computed by the layout autoencoder to represent the screen, as discussed in Section~\ref{sec:models}. With the \texttt{LayoutOnly} model, screens with similar layouts will have similar embedding vectors.

\tlhighlight{\paragraph{Visual Embedding Only} The \texttt{VisualOnly} model encodes the visual look of a screen by applying an autoencoder (described in Section~\ref{sec:models}) directly on its screenshot image bitmap instead of the layout bitmap. This baseline is inspired by the visual-based approach used in GUI task automation systems such as VASTA~\cite{sereshkeh2020vasta}, Sikuli~\cite{yeh_sikuli:_2009}, and HILC~\cite{intharah_hilc:_2019}. With the \texttt{VisualOnly} model, screens that are visually similar will have similar embedding vectors.}

\subsection{Prediction Task Results}
We report the performance on the GUI component and GUI screen prediction tasks of the \texttt{Screen2Vec} model, as well as the GUI screen prediction performance for the baseline models described above. 

Table~\ref{tab:gui_component_prediction_result} shows the top-1 accuracy (i.e., the top predicted GUI component matches the correct one), the top-0.01\% accuracy (i.e., the correct GUI component is among the top 0.01\% in the prediction result), the top-0.1\% accuracy, and the top-1\% accuracy of the two variations of the \texttt{Screen2Vec} model on the GUI component prediction task, where the model tries to predict the text content for each GUI component in all the GUI screens in the \textsc{Rico} dataset using its context (the other GUI components around it) among the collection of \textit{all} the GUI components in the \textsc{Rico} dataset.


\begin{table*}[htbp]
\small
\renewcommand{\arraystretch}{1.1}
\newcolumntype{L}{>{\raggedright\arraybackslash}X}
\begin{tabularx}{\textwidth}{|l|X|X|X|X|X|X|}

\hline
Model & Top-1 Accuracy & Top 0.01\% Accuracy & Top 0.1\% Accuracy & Top 1\% Accuracy & Top 5\% Accuracy & Top 10\% Accuracy\\
\hline
\texttt{Screen2Vec-EUCLIDEAN-text} & 0.443 & 0.619 & 0.783 & 0.856 & 0.885 & 0.901 \\
\hline
\tlhighlight{\texttt{Screen2Vec-HIERARCHICAL-text}} & 0.588 & 0.687 & 0.798 & 0.849 & 0.878 & 0.894 \\
\hline
\end{tabularx}
\vspace{0.3cm}
\caption{The GUI component prediction performance of the two variations of the \texttt{Screen2Vec} model with two different distance measures (\texttt{EUCLIDEAN} and \texttt{HIERARCHICAL}).}
\label{tab:gui_component_prediction_result}
\end{table*}


Similarly, Table~\ref{tab:gui_screen_prediction_result} reports the accuracy of the \texttt{Screen2Vec} model and the baseline models (\texttt{TextOnly}, \texttt{LayoutOnly}, and \texttt{VisualOnly}) on the task of predicting GUI screens, where each model tries to predict each GUI screen in all the GUI interaction traces in the \textsc{Rico} dataset using its context (the other GUI screens around it in the trace) among the collection of \textit{all} the GUI screens in the \textsc{Rico} dataset. For the \texttt{Screen2Vec} model, we compare three versions: one that encodes the locations of GUI components and the screen layouts and uses the \texttt{EUCLIDEAN} distance measure, one that uses such spatial information and the \texttt{HIERARCHICAL} distance measure, and one that uses the \texttt{EUCLIDEAN} distance measure without considering spatial information. A higher accuracy indicates that that the model is better at predicting the correct screen.

We also report the \textit{normalized} root mean square error (RMSE) of the predicted screen embedding vector for each model, normalized by the mean length of the actual screen embedding vectors. A smaller RMSE indicates that the top prediction screen generated by the model is, on average, more similar to the correct screen.

From the results in Table~\ref{tab:gui_screen_prediction_result}, we can see that the \texttt{Screen2Vec} models perform better than the baseline models in top-1 and top-k prediction accuracy. \tlhighlight{Among the different versions of \texttt{Screen2Vec}, the versions that encode locations of GUI components and the screen layouts performs better than the one without spatial information, suggesting that such spatial information is useful. The model that uses the \texttt{HIERARCHICAL} distance performs similarly to the one that uses the \texttt{EUCLIDEAN} distance in GUI component prediction, but performs worse in screen prediction.  In the \nameref{sec:downstream_tasks} section below, we will use the \texttt{Screen2Vec-\texttt{EUCLIDEAN}-spatial info} version of the \texttt{Screen2Vec} model.}   

\tlhighlight{As we can see, adding spatial information dramatically improves the Top-1 accuracy and the Top-0.01\% accuracy. However, the improvements in Top 0.1\% accuracy, Top 1\% accuracy, and normalized RMSE are smaller. We think the main reason is that aggregating the textual information, GUI class types, and app descriptions is useful for representing the high-level ``topic'' of a screen (e.g., a screen is about hotel booking because its text and app descriptions talk about hotels, cities, dates, rooms etc.), hence the good top 0.1\% and 1\% accuracy and normalized RMSE for the``no spatial info'' model. But these types of information are not sufficient for reliably differentiating the different types of screens needed (e.g., search, room details, order confirmation) in the hotel booking process because all these screens in the same app and task domain would contain ``semantically similar'' text. This is why the adding spatial information is helpful in identifying the top-1 and top-0.01\% results.}

Interestingly, the baseline models beat the ``no spatial info'' version of \texttt{Screen2Vec} in normalized RMSE: i.e., although the baseline models are less likely to predict \textit{the correct} screen, their predicted screens are, on average, more similar to the correct screen. A likely explanation to this phenomenon is that both baseline models use, by nature, similarity-based measures, while the \texttt{Screen2Vec} model is trained on a prediction-focused loss function. Therefore \texttt{Screen2Vec} does not emphasize making \textit{more similar} predictions when then prediction is incorrect. However, we can see that the \texttt{spatial info} versions of \texttt{Screen2Vec} perform better than the baseline models on both the prediction accuracy and the similarity measure.

\begin{table*}[htbp]
\small
\renewcommand{\arraystretch}{1.1}
\newcolumntype{L}{>{\raggedright\arraybackslash}X}
\begin{tabularx}{\textwidth}{|l|X|X|X|X|X|X|}
\hline
Model & Top-1 Accuracy & Top 0.01\% Accuracy & Top 0.1\% Accuracy & Top 1\% Accuracy & Top 5\% Accuracy & Normalized RMSE  \\
\hline
\texttt{Screen2Vec-EUCLIDEAN-spatial info} & 0.061 & 0.258 & 0.969 & 0.998 & 1.00 & 0.853 \\
\tlhighlight{\texttt{Screen2Vec-HIERARCHICAL-spatial info}} & 0.052 & 0.178 & 0.646 & 0.924 & 0.990 & 0.997 \\
\texttt{Screen2Vec-EUCLIDEAN-no spatial info} & 0.0065 & 0.116 & 0.896 & 0.986 & 0.999 & 1.723 \\

\texttt{TextOnly} & 0.012 & 0.055 & 0.196 & 0.439 & 0.643 &  1.241 \\
\texttt{LayoutOnly} & 0.0041 & 0.024 & 0.091 & 0.222 & 0.395 & 1.135 \\
\tlhighlight{\texttt{VisualOnly}} & 0.0060 & 0.026 & 0.121 & 0.252 & 0.603 & 1.543 \\

\hline
\end{tabularx}
\vspace{0.3cm}

\caption{The GUI screen prediction performance of the three variations of the \texttt{Screen2Vec} model and the baseline models (\texttt{TextOnly}, \texttt{LayoutOnly}, and \texttt{VisualOnly}).}
\label{tab:gui_screen_prediction_result}
\end{table*}

\tlcomment{TODO: fill in the numbers}

\tlcomment{Also test different configurations of Screen2Vec here}

\section{Sample Downstream Tasks}
\label{sec:downstream_tasks}
Note that while the accuracy measures are indicative of how much the model has learned about GUI screens and components, the main purpose of the \texttt{Screen2Vec} model is \textit{not} to predict GUI components or screens, but to produce distributed vector representations for them that encode useful semantic, layout, and design properties. Therefore this section presents several sample downstream tasks to illustrate important properties of the \texttt{Screen2Vec} representations and the usefulness of our approach.

\begin{figure*}
	\centering
	\includegraphics[width=0.85\textwidth]{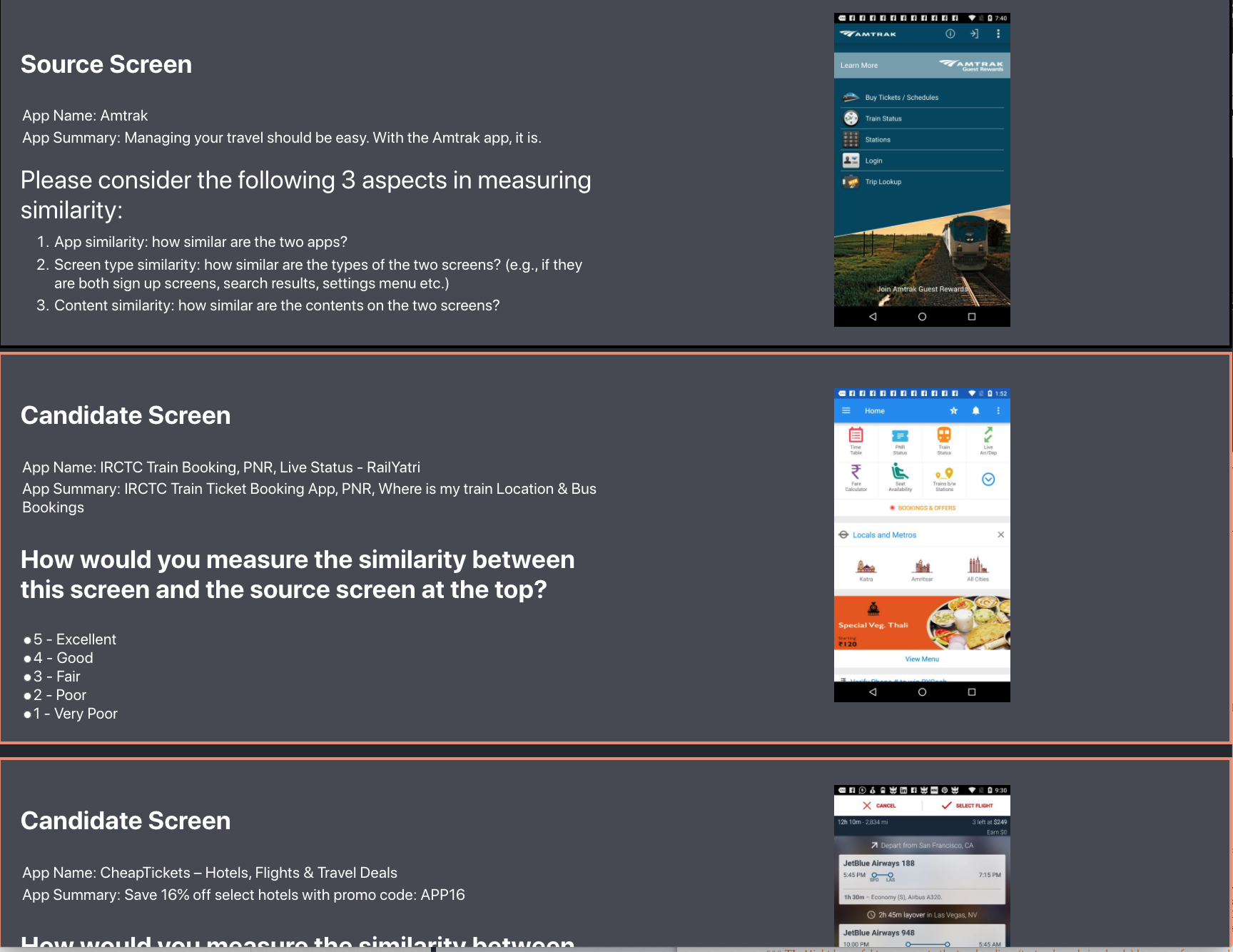}
    \Description{A web interface showing multiple-choice questions asking Mechanical Turk workers to rank the similarity between pairs of screens}
	\caption{The interface shown to the Mechanical Turk workers for rating the similarities for the nearest neighbor results generated by different models.}
	\label{fig:crowdui}
\end{figure*}

\subsection{Nearest Neighbors}
\label{sec:nearest_neighbors}

\tlcomment{Can we get useful nearest neighbors for GUI screens and components? What kinds of ``similarities'' get encoded in the result?}

The nearest neighbor task is useful for data-driven design, where the designers want to find examples for inspiration and for understanding the possible design solutions~\cite{deka_rico:_2017}. The task focuses on the similarity between GUI screen embeddings: for a given screen, what are the top-N most similar screens in the dataset? The similar technique can also be used for unsupervised clustering in the dataset to infer different types of GUI screens. In our context, this task also helps demonstrate the different characteristics between \texttt{Screen2Vec} and the three baseline models.

\tlcomment{Might be useful to compare to the two baselines (text-only and visual-only) here on a few examples}

We conducted a Mechanical Turk study to compare the similarity between the nearest neighbor results generated by the different models. We selected 50 screens from apps and app domains that most users are familiar with. We did \textit{not} select random apps from the \textsc{Rico} dataset, as many apps in the dataset would be obscure to Mechanical Turk workers so they might not understand them and therefore might not be able to judge the similarity of the results. For each screen, we retrieved the top-5 most similar screens using each of the 3 models. Therefore, each of the 50 screens had up to 3 (models) $\times$ 5 (screen each) = 15 similar screens, but many had fewer since different models may select the same screens. 

79 Mechanical Turk workers participated in this study\footnote{The protocol was approved by the IRB at our institution.}. In total, they labeled the similarity between 5,608 pairs of screens. Each worker was paid \$2 for each batch of 5 sets of source screens they labeled. A batch on average takes around 10 minutes to complete. In each batch, a worker went through a sample of 5 source screens from the 50 source screens in random order, where for each source screen, the worker saw the union of the top-5 most similar screens to the source screen generated by the 3 models in random order. For each screen, we also showed the worker the app it came from and a short description of the app from the Google Play Store, but we did not show them which model produced the screen. The worker was asked to rate the similarity of each screen to the original source screen on a scale of 1 to 5 (Figure~\ref{fig:crowdui}). We asked the workers to consider 3 aspects in measuring similarity: (1) app similarity (how similar are the two apps); (2) screen type similarity (how similar are the types of the two screens e.g., if they are both sign up screens, search results, settings menu etc.); and (3) content similarity (how similar are the content on the two screens). 

Table~\ref{tab:crowd_result} shows the mean screen similarity rated by the Mechanical Turk workers for the top-5 nearest neighbor results of the sample source screens generated by the 3 models. The Mechanical Turk workers rated the nearest neighbor screens generated by the \texttt{Screen2Vec} model to be, on average, more similar to their source screens than the nearest neighbor screens generated by the baseline \texttt{TextOnly} and \texttt{LayoutOnly} models. \tlhighlight{Tested with a non-parametric Mann-Whitney U test (because the ratings are not normally distributed), the differences between the mean ratings of the \texttt{Screen2Vec} model and both the \texttt{TextOnly} model and the \texttt{LayoutOnly} model are significant ($p<0.0001$).}

\begin{table*}[htbp]
    \small
    \renewcommand{\arraystretch}{1.1}
    \newcolumntype{L}{>{\raggedright\arraybackslash}X}

    \begin{tabularx}{\textwidth}{|X|X|X|X|X|X|}
        \hline
        \multicolumn{2}{|c|}{\texttt{Screen2Vec}} & \multicolumn{2}{|c|}{\texttt{TextOnly}} & \multicolumn{2}{|c|}{\texttt{LayoutOnly}}   \\
        \hline
        Mean Rating & Std. Dev. & Mean Rating & Std. Dev. & Mean Rating & Std. Dev. \\
        \hline
        3.295* & 1.238 & 3.014* & 1.321 & 2.410* & 1.360 \\
        \hline
    \end{tabularx}

    \vspace{0.3cm}

    \caption{The mean screen similarity rated by the Mechanical Turk workers for the top-5 nearest neighbor results of the sample source screens generated by the 3 models: \texttt{Screen2Vec}, \texttt{TextOnly}, and \texttt{LayoutOnly} (\textit{*p<0.0001}).}
    \label{tab:crowd_result}
\end{table*}


\begin{figure*}
	\centering
	\includegraphics[width=\textwidth]{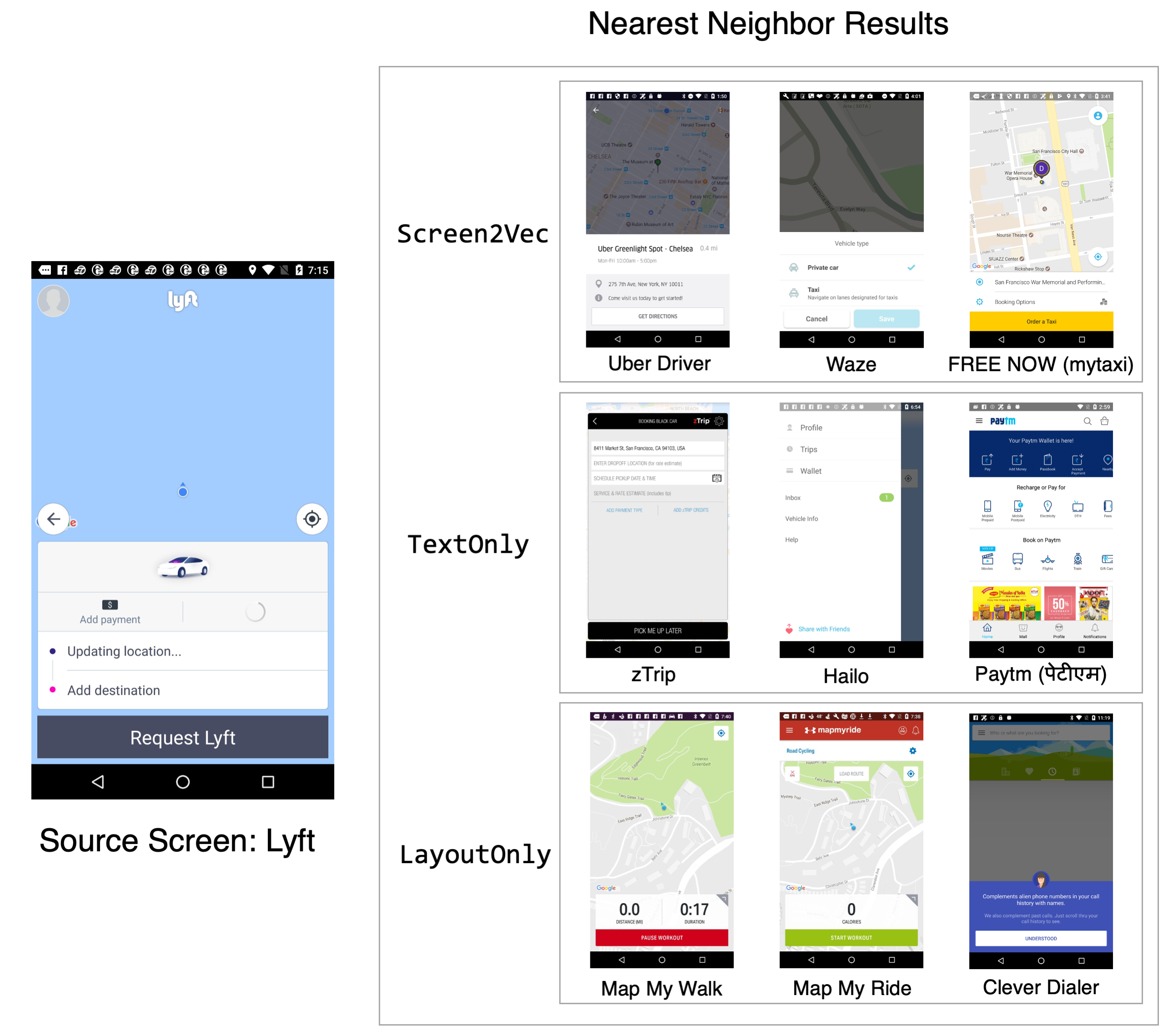}
	\Description{Screen2Vec generates the nearest neighbor screens for the ``request ride'' screen of the Lyft app.}
	\caption{The example nearest neighbor results for the Lyft ``request ride'' screen generated by the \texttt{Screen2Vec}, \texttt{TextOnly}, and \texttt{LayoutOnly} models.}
	\label{fig:nearestneighborexample}
\end{figure*}

Subjectively, when looking at the nearest neighbor results, we can see the different aspects of the GUI screens that each different model captures. \texttt{Screen2Vec} can create more comprehensive representations that encode the textual content, visual design and layout patterns, and app contexts of the screen compared with the baseline models, which only capture one or two aspects. For example, Figure~\ref{fig:nearestneighborexample} shows the example nearest neighbor results for the ``request ride'' screen in the Lyft app. \texttt{Screen2Vec} model retrives the ``get direction'' screen in the Uber Driver app, ``select navigation type'' screen in the Waze app, and ``request ride'' screen in the Free Now (My Taxi) app. Considering the Visual and component layout aspects, the result screens all feature a menu/information card at the bottom 1/3 to 1/4 of the screen, with a \texttt{MapView} taking the majority of the screen space. Considering the content and app domain aspects, all of these screens are from transportation-related apps that allow the user to configure a trip. In comparison, the \texttt{TextOnly} model retrieves the ``request ride'' screen from the zTrip app, the ``main menu'' screen from the Hailo app (both zTrip and Hailo are taxi hailing apps), and the home screen of the Paytm app (a mobile payment app in India). The commonality of these screens is that they all include text strings that are semantically similar to ``payment'' (e.g., add payment type, wallet, pay, add money), and strings that are semantically similar to ``destination'' and ``trips'' (e.g., drop off location, trips, bus, flights). But the model did not consider the visual layout and design patterns of the screens nor the app context. Therefore the result contains the ``main menu'' (a quite different type of screen) in the Hailo app and the ``home screen'' in the Paytm app (a quite different type of screen in a different type of app). The \texttt{LayoutOnly} model, on the other hand, retrieves the ``exercise logging'' screens from the Map My Walk app and the Map My Ride app, and the tutorial screen from the Clever Dialer app. We can see that the content and app-context similarity of the result of the \texttt{LayoutOnly} model is quite lower than those of the \texttt{Screen2Vec} and \texttt{TextOnly} models. However, the result screens all share similar layout features as the source screen, such as the menu/information card at the bottom of the screen and the screen-wide button at the bottom of the menu. 


\begin{figure*}
	\centering
	\includegraphics[width=\textwidth]{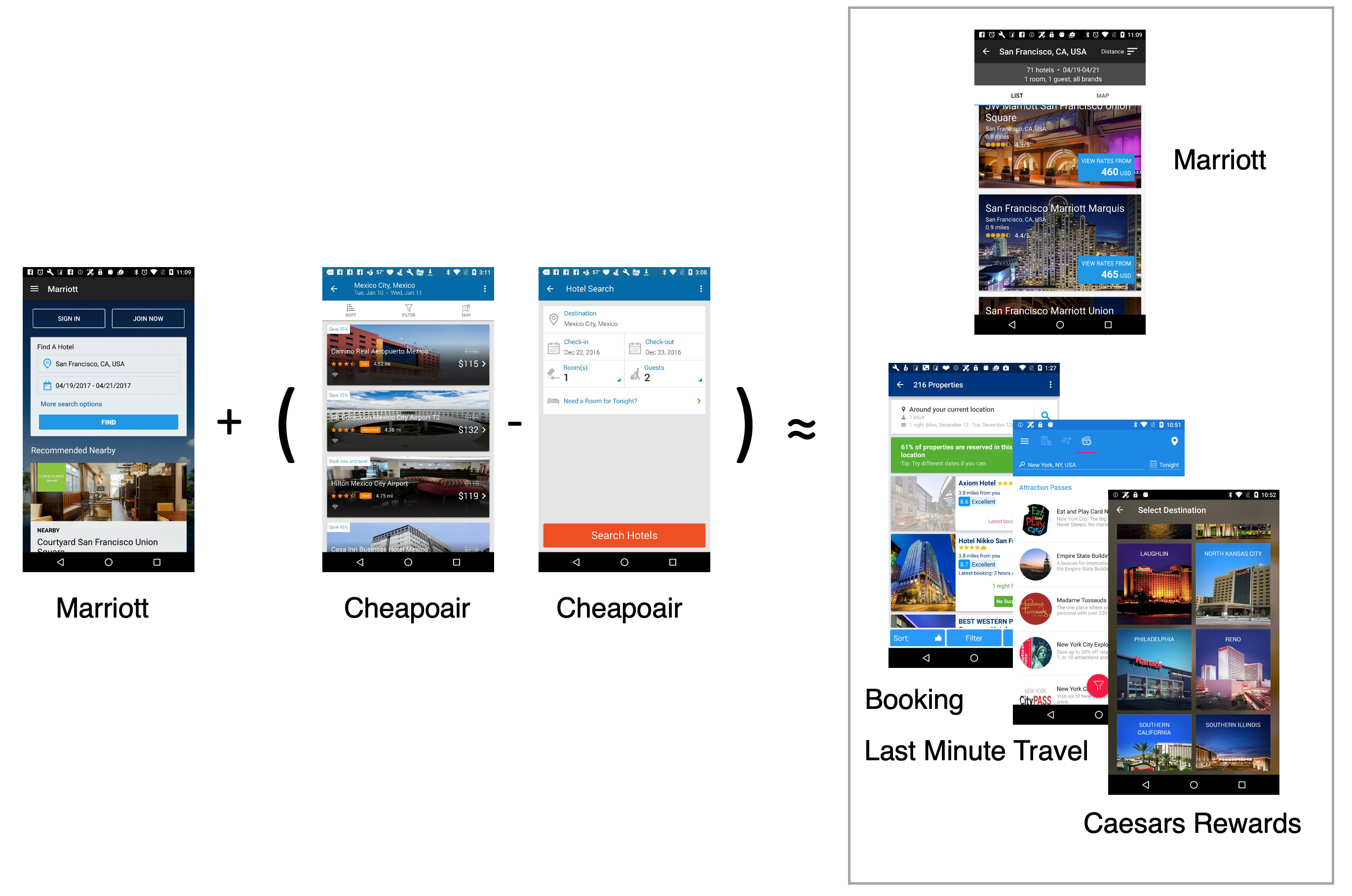}
	\Description{An expression showing that adding the hotel booking page of the Marriott app to the search results page of the Cheapoair app, and substracting the hotel booking page of the Cheapoair app can result in the search result page in the Marriott app and the similar pages of a few other travel apps.}
	\caption{An example showing the composability of \texttt{Screen2Vec} embeddings: running the nearest neighbor query on the composite embedding of Marriott app 's hotel booking page $+$ Cheapoair app's hotel booking page $-$ Cheapoair app's search result page can match the Marriott app's search result page and the similar pages of a few other travel apps.} 
	\label{fig:composability_figure}
\end{figure*}

\subsection{Embedding Composability}
\label{sec:embedding_composability}

 \tlcomment{Does the representations exhibit useful additive composability? e.g., Can we get the McDonald's checkout page = McDonald's order page + (Starbucks checkout page - Starbucks order page)?}

A useful property of embeddings is that they are composable---meaning that we can add, subtract, and average embeddings to form a meaningful new one. This property is commonly used in word embeddings. For example, in \texttt{Word2Vec}, analogies such as ``man is to woman as brother is to sister'' is reflected in that the vector $(man-woman)$ is similar to the vector $(brother-sister)$. Besides representing analogies, this embedding composability can also be utilized for generative purposes---for example, $(brother-man+woman)$ results in an embedding vector that represents ``sister''.

This property is also useful in screen embeddings. For example, we can run a nearest neighbor query on the composite embedding of (Marriott app 's ``hotel booking'' screen $+$ (Cheapoair app's ``search result'' screen $-$ Cheapoair app's ``hotel booking'' screen)). The top result is the ``search result'' screen in the Marriott app (see Figure~\ref{fig:composability_figure}). When we filter the result to focus on screens from apps other than Marriott, we get screens that show list results of items from other travel-related mobile apps such as Booking, Last Minute Travel, and Caesars Rewards.

The composability can make \texttt{Screen2Vec} particularly useful for GUI design purposes---the designer can leverage the composability to find inspiring examples of GUI designs and layouts. We will discuss more about its potential applications in Section~\ref{sec:application}.

\subsection{Screen Embedding Sequences for Representing Mobile Tasks}
\label{sec:embedding_tasks}
\tlcomment{Can we combine embeddings of GUI screens / invoked GUI components to represent a sequence of GUI actions that correspond to a task?}

\tlcomment{Can we use such representations of tasks to estimate task similarity?}

\tlcomment{Can we use such representation of tasks to transfer learned procedures? E.g., what's the equivalent of clicking on GUI component E1 on screen S1 for task T1 when on a different screen S2?}

GUI screens are not only useful data sources individually on their own, but also as building blocks to represent a user's task. A task in an app, or across multiple apps, can be represented as a sequence of GUI screens that makes up the user interaction trace for performing this task using app GUIs. In this section, we conduct a preliminary evaluation on the effectiveness of embedding mobile tasks as sequences of \texttt{Screen2Vec} screen embedding vectors.

Similar to GUI screens and components, the goal of embedding mobile tasks is to represent them in a vector space where more similar tasks are closer to each other. To test this, we recorded the scripts of completing 10 common smartphone tasks, each with two variations that use different apps, using our open-sourced \textsc{Sugilite}~\cite{li_sugilite:_2017} system on a Pixel 2 XL phone running Android 8.0. Each script consists of a sequence of ``perform action X (e.g., click, long click) on the GUI component Y in the GUI screen Z''. In this preliminary evaluation, we only used the screen context: we represented each task as the average of the \texttt{Screen2Vec} screen embedding vectors for all the screens in the task sequence.  

Table~\ref{tab:task_embedding_tasks} shows the 10 tasks we tested on, the two apps used for each task, and the number of unique GUI screens in each trace used for task embedding. We queried for the nearest neighbor within the 20 task variations for each task variation, and checked if the model could correctly identify the similar task that used a different app. The \texttt{Screen2Vec} model achieved a 18/20 (90\%) accuracy in this test.  In comparison, when we used the \texttt{TextOnly} model for task embedding, the accuracy was 14/20 (70\%). 

\begin{table*}[htbp]
    \small
    \renewcommand{\arraystretch}{1.1}
    \newcolumntype{L}{>{\raggedright\arraybackslash}X}
    \begin{tabularx}{\textwidth}{|l|X|X|X|X|}
        \hline
        Task Description & App 1 & Screen Count & App 2 & Screen Count \\
        \hline
        Request a cab & Lyft & 3 & Uber & 2 \\
        Book a flight & Fly Delta & 4 & United Airlines & 4 \\
        Make a hotel reservation & Booking.com & 7 & Expedia & 7 \\
        Buy a movie ticket & AMC Theaters & 3 & Cinemark & 4 \\
        Check the account balance & Chase & 4 & American Express & 3 \\
        Check sports scores & ESPN & 4 & Yahoo! Sports & 4 \\
        Look up the hourly weather & AccuWeather & 3 & Yahoo! Weather & 3 \\
        Find a restaurant & Yelp & 3 & Zagat & 4 \\
        Order an iced coffee & Starbucks & 7 & Dunkin' Donuts & 8 \\
        Order takeout food & GrubHub & 4 & Uber Eats & 3 \\
        \hline
        
    \end{tabularx}
    \vspace{0.3cm}

    \caption{A list of 10 tasks we used for the preliminary evaluation of using \texttt{Screen2Vec} for task embedding, along with the apps used and the count of screens used in the task embedding for each variation.}
    \label{tab:task_embedding_tasks}
\end{table*}

While the task embedding method we explored in this section is quite primitive, it illustrates that the \texttt{Screen2Vec} technique can be used to effectively encode mobile tasks into the vector space where semantically similar tasks are close to each other. For the next steps, we plan to further explore this direction. For example, the current method of averaging all the screen embedding vectors does not consider the order of the screens in the sequence. In the future, we may collect a dataset of human annotations of task similarity, and use techniques that can encode the sequences of items, such as recurrent neural networks (RNN) and long short-term memory (LSTM) networks, to create the task embeddings from sequences of screen embeddings. We may also incorporate the \texttt{Screen2Vec} embeddings of the GUI components that were interacted with (e.g., the button that was clicked on) to initiate the screen change into the pipeline for embedding tasks. 

\tlcomment{limitation: did not consider the target component, did not consider the order of the sequence at all etc.}

\section{Potential Applications}
\label{sec:application}
This section describes several potential applications where the new \texttt{Screen2Vec} technique can be useful based on the downstream tasks described in Section~\ref{sec:downstream_tasks}.

\texttt{Screen2Vec} can enable new GUI design aids that take advantage of the nearest neighbor similarity and composability of \texttt{Screen2Vec} embeddings. Prior work~\cite{deka_rico:_2017,kumar_webzeitgeist:_2013,huang_swire_2019} has shown that data-driven tools that enable designers to curate design examples are useful for interface designers. Unlike~\cite{deka_rico:_2017}, which uses a content-agnostic approach that focuses on the visual and layout similarities, \texttt{Screen2Vec} considers the textual content and app meta-data in addition to the visual and layout patterns, often leading to different nearest neighbor results as discussed in Section~\ref{sec:nearest_neighbors}. This new type of similarity results will also be useful when focusing on interface design beyond just visual and layout issues, as the results enable designers to query for example designs that display similar content or screens that are used in apps in a similar domain. The composability in {Screen2Vec} embeddings enables querying for design examples at a finer granularity. For example, suppose a designer wishes to find examples for inspiring the design of a new checkout page for app A. They may query for the nearest neighbors of the 
synthesized embedding App A's order page $+$ (App B's checkout page $-$ App B's order page). Compared with only querying for the nearest neighbors of App B's checkout page, this synthesized query encodes the interaction context (i.e., the desired page should be the checkout page for App A's order page) in addition to the ``checkout'' semantics.

The \texttt{Screen2Vec} embeddings can also be useful in generative GUI models. Recent models such as the neural design network (NDN)~\cite{lee_neural_2020} and LayoutGAN~\cite{li_layoutgan_2019} can generate realistic GUI layouts based on user-specified constraints (e.g., alignments, relative positions between GUI components). \texttt{Screen2Vec} can be used in these generative approaches to incorporate the semantics of GUIs and the contexts of how each GUI screen and component gets used in user interactions. For example, the GUI component prediction model can estimate the likelihood of each GUI component given the context of the other components in a generated screen, providing a heuristic of how likely the GUI components would fit well with each other. Similarly, the GUI screen prediction model may be used as a heuristic to synthesize GUI screens that would better fit with the other screens in the planned user interaction flows. Since \texttt{Screen2Vec} has been shown effective in representing mobile tasks in Section~\ref{sec:embedding_tasks}, where similar tasks will yield similar embeddings, one may also use the task embeddings of performing the same task on an existing app to inform the generation of new screen designs. The embedding vector form of \texttt{Screen2Vec} representations would make them particularly suitable for use in the recent neural-network based generative models. 

\texttt{Screen2Vec}'s capability of embedding tasks can also enhance interactive task learning systems. Specifically, \texttt{Screen2Vec} may be used to enable more powerful procedure generalizations of the learned tasks. We have shown that the \texttt{Screen2Vec} model can effectively predict screens in an interaction trace. Results in Section~\ref{sec:embedding_tasks} also indicated that \texttt{Screen2Vec} can embed mobile tasks so that the interaction traces of completing the same task in different apps will be similar to each other in the embedding vector space. Therefore, it is quite promising that \texttt{Screen2Vec} may be used to generalize a task learned from the user by demonstration in one app to another app in the same domain (e.g., generalizing the procedure of ordering coffee in the Starbucks app to the Dunkin' Donut app). In the future, we plan to further explore this direction by incorporating \texttt{Screen2Vec} into open-sourced mobile interactive task learning agents such as our \textsc{Sugilite} system~\cite{li_sugilite:_2017}.





\section{Limitations and Future Work}
There are several limitations of our work in \texttt{Screen2Vec}. First, \texttt{Screen2Vec} has only been trained and tested on Android app GUIs. \tlhighlight{However, the approach used in \texttt{Screen2Vec} should apply to any GUI-based apps with hierarchical-based structures (e.g., view hierarchies in iOS apps and hierarchical DOM structures in web apps). We expect embedding desktop GUIs to be more difficult than mobile ones, because individual screens in desktop GUIs are usually more complex with more heterogeneous design and layout patterns.} 

Second, the \textsc{Rico} dataset we use only contains interaction traces within single apps. The approach used in \texttt{Screen2Vec} should generalize to interaction traces across multiple apps. We plan to evaluate its prediction performance on cross-app traces in the future with an expanded dataset of GUI interaction traces. The \textsc{Rico} dataset also does not contain screens from paid apps, screens that require special accounts/privileges to access to (screens that require free accounts to access are included when the account registration is readily available in the app), or screens that require special hardware (e.g., in the companion apps for smart home devices) or specific context (e.g., pages that are only shown during events) to access. This limitation of the \textsc{Rico} dataset might affect the performance of the pre-trained \texttt{Screen2Vec} model on these underrepresented types of app screens.

A third limitation is that the current version of \texttt{Screen2Vec}
does not encode the semantics of graphic icons that have no textual information. Accessibility-compliant apps all have alternative texts for their graphic icons, which \texttt{Screen2Vec} already encodes in its GUI screen and component embeddings as a part of the text embedding. However, for non-accessible apps, computer vision-based (e.g.,~\cite{chen2020unblind, liu2018learning}) or crowd-based (e.g.,~\cite{zhang_interaction_2017}) techniques can be helpful for generating textual annotations for graphic icons so that their semantics can be represented in \texttt{Screen2Vec}. \tlhighlight{Another potentially useful kind of information is the rules and examples in GUI design systems (e.g., Android Material Design, iOS Design Patterns). While Screen2Vec can, in some ways, ``learn'' these patterns from the training data, it will be interesting to explore a hybrid approach that can leverage their explicit notions. We will explore incorporating these techniques into the \texttt{Screen2Vec} pipeline in the future.}

\balance

\section{Related Work}
\subsection{Distributed Representations of Natural Language}
The study of representing words, phrases, and documents as mathematical objects, often vectors, is central to natural language processing (NLP) research~\cite{turian_word_2010, mikolov_distributed_2013}. Conventional non-distributed word embedding methods represent a word using a \textit{one-hot} representation where the vector length equals the size of the vocabulary, and only one dimension (that corresponds to the word) is on~\cite{turian_word_2010}. This representation does not encode the semantics of the words, as the vector for each word is perpendicular to the others. Documents represented using a one-hot word representation also suffer from the curse of dimensionality~\cite{bellman-dynamic-1966} as a result of the extreme sparsity in the representation.

By contrast, a \textit{distributed} representation of a word represents the word across multiple dimensions in a continuous-valued vector (word embedding)~\cite{bengio2009learning}. Such distributed representations can capture useful syntactic and semantic properties of the words, where syntactically and semantically related words are similar in this vector space~\cite{turian_word_2010}. Modern word embedding approaches usually use the language modeling task. For example, \texttt{Word2Vec}~\cite{mikolov_distributed_2013} learns the embedding of a word by predicting it based on its context (i.e., surrounding words), or predicting the context of a word given the word itself. GloVe~\cite{pennington-etal-2014-glove} is similar to \texttt{Word2Vec} on a high level, but focuses on the likelihood that each word appears in the context of other words with in the whole corpus of texts, as opposed to \texttt{Word2Vec} which uses local contexts. More recent work such as ELMo~\cite{peters-etal-2018-deep} and BERT~\cite{devlin_2018_bert} allowed contextualized embedding. That is, the representation of a phrase can vary depending on a word's context to handle polysemy (i.e., the capacity for a word or phrase to have multiple meanings). For example, the word ``bank'' can have different meanings in ``he withdrew money from the bank'' versus ``the river bank''

While distributed representations are commonly used in natural language processing, to our best knowledge, the \texttt{Screen2Vec} approach presented in this paper is the first to seek to encode the semantics, the contexts, and the design patterns of GUI screens and components using distributed representations. The \texttt{Screen2Vec} approach is conceptually similar to \texttt{Word2Vec} on a high level---like \texttt{Word2Vec}, \texttt{Screen2Vec} is trained using a predictive modeling task where the context of a target entity (words in \texttt{Word2Vec}, GUI components and screens in \texttt{Screen2Vec}) is used to predict the entity (known as the continuous bag of words (CBOW) model in \texttt{Word2Vec}). There are also other relevant \texttt{Word2Vec}-like approaches for embedding APIs  based their usage in source code and software documentations (e.g., \texttt{API2Vec}~\cite{nguyen_exploring:_2017}), and modeling the relationships between user tasks, system commands, and natural language descriptions in the same vector space (e.g., CommandSpace~\cite{adar_commandspace:_2014}).

Besides the domain difference between our \texttt{Screen2Vec} model and \texttt{Word2Vec} and its follow-up work, \texttt{Screen2Vec} uses both a (pre-trained) text embedding vector and a class type vector, and combines them with a linear layer. It also incorporates external app-specific meta-data such as the app store description. The hierarchical approach allows \texttt{Screen2Vec} to compute a screen embedding with the embeddings of the screen's GUI components, as described in Section~\ref{sec:approach}. In comparison, \texttt{Word2Vec} only computes word embeddings using word contexts without using any other meta-data~\cite{mikolov_distributed_2013}.

\subsection{Modeling GUI Interactions}
\texttt{Screen2Vec} is related to prior research on computationally modeling app GUIs and the GUI interactions of users. The interaction mining approach~\cite{deka_erica:_2016} captures the static (UI layout, visual features) and dynamic (user flows) parts of an app's design from a large corpus of user interaction traces with mobile apps, identifies 23 common flow types (e.g., adding, searching, composing), and can classify the user's GUI interactions into these flow types. A similar approach was also used to learn the design semantics of mobile apps, classifying GUI elements into 25 types of GUI components, 197 types of text buttons, and 135 types of icon classes~\cite{liu2018learning}. Appstract~\cite{fernandes_appstract:_2016} focused on the semantic entities (e.g., music, movie, places) instead, extracting entities, their properties, and relevant actions from mobile app GUIs. These approaches use a smaller number of discrete 
types of flows, GUI elements, and entities to represent GUI screens and their components, while our \texttt{Screen2Vec} uses continuous embedding in a vector space for screen representation. 

Some prior techniques specifically focus on the visual aspect of GUIs. The \textsc{Rico} dataset~\cite{deka_rico:_2017} shows that it is feasible to train a GUI layout embedding with a large screen corpus, and retrieve screens with similar layouts using such embeddings. Chen et al.'s work~\cite{chen2020unblind} and Li et al.'s work~\cite{li-etal-2020-widget} show that a model can predict semantically meaningful alt-text labels for GUI components based on their visual icon. \texttt{Screen2Vec} provides a more holistic representation of GUI screens by encoding textual content, GUI component class types, and app-specific meta-data in addition to the visual layout.

Another category of work in this area focuses on predicting GUI actions for completing a task objective. Pasupat et al.'s work~\cite{pasupat-etal-2018-mapping} maps the user's natural language commands to target elements on web GUIs. Li et al.'s work~\cite{li_mapping:_2020} goes a step further by generating sequences of actions based on natural language commands. These works use a supervised approach that requires a large amount of manually-annotated training data, which limits its utilization. In comparison, \texttt{Screen2Vec} uses a self-supervised approach that does not require any manual data annotation of user intents and tasks. \texttt{Screen2Vec} also does not require any annotations of the GUI screens themselves, unlike~\cite{zhang_robust_2018} which requires additional developer annotations as meta-data for GUI components.

\subsection{Interactive Task Learning}
Understanding and representing GUIs is a central challenge in GUI-based interactive task learning (ITL). When the user demonstrates a task in an app, the system needs to understand the user's action in the context of the underlying app GUIs so that it can generalize what it has learned to future task contexts~\cite{li_appinite:_2018}. For example, \textsc{Sugilite} represents each app screen as a graph where each GUI component is an entity~\cite{li_interactive:_2020}. Properties of GUI components, their hierarchical relations, and the spatial layouts are represented as edges in the graph. This graph representation allows grounding natural language instructions to GUIs~\cite{li_appinite:_2018,li_interactive:_2020} with graph queries, allowing a more natural end user development experience~\cite{myers_making_2017}. It also supports personal information anonymization on GUIs~\cite{li_privacy:_2020}. However, this graph representation is difficult to aggregate or compare across different screens or apps. Its structure also does not easily fit into common machine learning techniques for computationally modeling the GUI tasks. As a result, the procedure generalization capability of systems like \textsc{Sugilite} is limited to parameters within the same app and the same set of screens.

Some other interactive task learning systems such as \textsc{Vasta}~\cite{sereshkeh2020vasta}, Sikuli~\cite{yeh_sikuli:_2009}, and \textsc{Hilc}~\cite{intharah_hilc:_2019} represent GUI screens visually. This approach performs segmentation and classification on the video of the user performing GUI actions to extract visual representations (e.g., screenshot segments/icons) of GUI components, allowing replay of actions by identifying target GUI components using computer vision object recognition techniques. This approach supports generalization based on visual similarity (e.g., perform an action on \textit{all} PDF files in a file explorer because they all have visually similar icons). However, this visual approach is limited by its lack of semantic understanding of the GUI components. For example, the icon of a full trash bin is quite different from an that of an empty one pixel count wise, but they should have the same meaning when the user intent is ``open the trash bin''. The icon for a video file can be similar to that of an audio file (with the only difference being the tiny ``mp3`` and ``mp4`` at a corner), but the system should differentiate them in intents like ``select all the video files''. 

The \texttt{Screen2Vec} representation presented in this paper encodes the textual content, visual layout and design patterns, and app-specific context of GUI screens in a distributed vector form that can be used across different apps and task domains. We think this representation can be quite useful in \textit{supplementing} the existing graph and visual GUI representations in ITL systems. For example, as shown in Section~\ref{sec:embedding_tasks}, sequences of \texttt{Screen2Vec} screen embedding can represent tasks in a way that allows the comparison and retrieval of similar tasks among different apps. The results in Section~\ref{sec:embedding_tasks} also suggest that the embedding can help an ITL agent transfer procedures learned from one app to another. 



\section{Conclusion}

We have presented \texttt{Screen2Vec}, a new self-supervised technique for generating distributed semantic representations of GUI screens and components using their textual content, visual design and layout patterns, and app meta-data. This new technique has been shown to be effective in downstream tasks such as nearest neighbor retrieval, composability-based retrieval, and representing mobile tasks. \texttt{Screen2Vec} addresses an important gap in computational HCI research, and could be utilized for enabling and enhancing interactive systems in task learning (e.g.,~\cite{li_pumice:_2019, sereshkeh2020vasta}), intelligent suggestive interfaces (e.g.,~\cite{Chen2019MessageOnTap}), assistive tools (e.g.,~\cite{bigham2009trailblazer}), and GUI design aids (e.g.,~\cite{swearngin2018rewire,lee2020guicomp}).

\begin{acks}
This research was supported in part by Verizon through the Yahoo! InMind project, a J.P. Morgan Faculty Research Award, Google Cloud Research Credits, NSF grant IIS-1814472, and AFOSR grant FA95501710218. Any opinions, findings or recommendations expressed here are those of the authors and do not necessarily reflect views of the sponsors. We would like to thank our anonymous reviewers for their feedback and Ting-Hao (Kenneth) Huang, Monica Lam, Vanessa Hu, Michael Xieyang Liu, Haojian Jin, and Franklin Mingzhe Li for useful discussions.
\end{acks}

\vspace*{\fill}
\bibliographystyle{ACM-Reference-Format}
\bibliography{references}


\end{document}